\documentstyle[prd,aps,preprint]{revtex}
\tightenlines
\newcommand{\los}{\raisebox{-0.6ex}{$\scriptstyle
\buildrel{\raisebox{-1pt}{$<$}}\over\sim$}} % less or similar...
\title{On the potentials yielding cosmological scaling solutions}
\author{Ana Nunes$^{1,2}$\thanks{email: anunes@lmc.fc.ul.pt} \& Jos\'e P. Mimoso$^{1,3}$\thanks{email:jpmimoso@alf1.cii.fc.ul.pt}
}
\address{$^{1}$Dep. F\'{\i}sica, Faculdade de Ci\^encias, Ed C8, Campo Grande\\
Universidade de Lisboa,  1700
Lisboa, Portugal}
\address{$^2$CMAF and $^3$CFNUL, Av. Prof. Gama Pinto, 2, 1699 Lisboa Codex, Portugal}
\date{\today}
\begin{document}
%\preprint{}
%
\maketitle
\begin{abstract}
In the present work we perform a phase-plane analysis of the complete dynamical system corresponding to a flat FRW cosmological models with a perfect fluid and a self-interacting scalar field and show that every
positive and monotonous potential which is asymptotically exponential yields a scaling solution as a global attractor.
\end{abstract}

%%%%%%%%%%%%%%
%\pacs{98.80.Hw, 98.80.Cq, 04.50.+h, 04.60.+n} %Need to be updated
%%%%%%%%%%%%%%%%%%%%%%%%%%%%%%%%%%%%%%%%%%%%%%%%%%%%%%%%%%%%%%%%%%%%%%

%%%%%%%%%%
\section{Introduction}
Scaling solutions of cosmological models with a self-interacting
scalar field have attracted a lot of interest recently. It has
been advocated that scalar field cosmologies offer a way to
reconcile the standard cosmological model with
the large scale structure (LSS) data~\cite{White et al 93} and the remarkable results of
the supernovae observations~\cite{Perlmut et al 98} that indicate that the baryonic matter
density parameter is bound to be $\Omega_m \los 0.3$ and that the
universe is positively accelerated at present. Scalar fields
would thus act like a decaying cosmological constant, but would otherwise
have the advantage of evading some of the longstanding
difficulties faced by the latter~\cite{Coble et al 97,Viana+Liddle 98}.

Rather than knowing the exact solutions of the scalar field cosmologies, 
what matters most is the asymptotic behaviour of models. The self-similar 
solutions that have been found when the potential of the self-interacting
scalar field is exponential are particular remarkable in this sense~\cite{Wetterich 88,WCL 93,Ferreira+Joyce 97,CLW 98}. 
They correspond to an equilibrium between the different kinds of energy 
that compete. Indeed the energy density of the scalar field scales 
with the same rate as that of the barotropic matter (some authors~\cite{Liddle+Scherrer 99} refer to these solutions as {\em trackers} and denote {\em scaling} solutions those for which the energy density of the scalar field evolves as a power of that of matter; in what follows we shall adopt this terminology). Moreover
these solutions attract all the other phase space trajectories in the case of flat space models
and, hence, provide the late-time asymptotic behaviour for the 
scalar field cosmologies under consideration. This gives, on the 
one hand, a possible answer for why a non-vanishing scalar field 
does not introduce radical changes with respect to 
the usual Einstein-de Sitter rate of expansion of the universe. 
On other hand, it may contribute an explanation to the 
difference between the actual  density of matter and the critical 
energy density of the spatially flat isotropic models. Futhermore, the scalar field component would also fulfill the convenient role of delaying
the time of matter-radiation equality which would help fitting
the power spectrum of large scale structure~\cite{Coble et al 97,Wetterich 95,Viana+Liddle 98}.

In the literature we find mainly two sorts of potentials underlying the tracking and the scaling behaviour: the exponential potentials and a class of power law potentials~\cite{Ratra+Peebles 88,Liddle+Scherrer 99,Zlatev+Wang+Steinhardt 99} respectively. However it is worth noticing that the solutions corresponding to the latter set of potentials were only shown to hold in the regime where the perfect fluid
component fully dominates the expansion and the energy density of the scalar field is negligible. Regarding
the solutions termed trackers~\cite{Liddle+Scherrer 99}: the self-adjustment of the behaviour of the energy density of one or more scalar fields with that of matter has been investigated in Friedmann-Robertson-Walker (FRW) models~\cite{Wetterich 88,WCL 93,Ferreira+Joyce 97,CLW 98,Billyard+Coley+Hoogen 98,Hoogen+Coley+Wands 99} both with and without curvature, in spatially homogeneous, but anisotropic models~\cite{Coley+Ibanez+Hoogen 97}, and in FRW models in scalar-tensor gravity theories (also refered to as non-minimal coupling)~\cite{Billyard+Coley+Ibanez 98,Uzan 99,Amendola 99,Holden+Wands 00}.  

Here we address the question of whether there are any classes of potentials, besides the exponential, yielding the self-similar behaviour found in the latter case. In general the motivation for considering exponential potentials is drawn from supergravity (superstrings) where these type of potentials commonly arise after the dimensional reduction into a 4-dimensional space-time theory. However the resulting effective potentials are seldom simply exponentials and exhibit some form factors due to couplings 
between fields. On other hand, it has been claimed that the only potentials inducing the tracker scaling solutions are the exponential potentials~\cite{Liddle+Scherrer 99}. In the present work we perform a phase-plane analysis of the complete dynamical system corresponding to a flat FRW cosmological models with a perfect fluid and a self-interacting scalar field and show that every
positive and monotonous potential which is asymptotically exponential also yields a tracking solution. Our results are the outcome of taking into consideration the full third order dynamical system. In previous work in the literature the analysis which were carried out miss an equation for the scalar field and the resulting reduced system is only appropriate to study the exponential potential cases. 

\section{Scalar field cosmologies}
Consider the flat homogeneous and isotropic universes given by the
Friedmann-Robertson-Walker (FRW) metric
\begin{equation}
{\rm d}s^2 = - {\rm d}t^2 + a^2(t)\,\left[{\rm
d}r^2+r^2({\rm d}\theta^2+\sin^2\theta\,{\rm d}\phi^2
\right] \; . \label{Fried_met}
\end{equation}
Assume that the matter sources are a perfect
fluid characterized by the equation of state $p=(\gamma
-1)\,\rho$, where $0\le \gamma\le 2$ is a constant, and a
self-interacting scalar field $\varphi$ with the potential
$V(\varphi)$. The field equations then read
\begin{eqnarray}
H^2 &=& \frac{8\pi G}{3}\,
\left(\frac{\dot{\varphi}^2}{2} + V(\varphi) + \rho \right)
\label{e:Fried}\; , \\
\ddot{\varphi}+3\frac{\dot{a}}{a}\,\dot{\varphi}&=&
-\frac{\partial V(\varphi)}{\partial \varphi} \label{e:sf1}\; ,
\end{eqnarray}
where the overdots stand for the derivatives with respect to the
time $t$, and $H=\dot{a}/a$. Another equation which is useful is

\begin{equation}
\dot{H} = - \frac{\aleph}{2}\,\left(\dot{\varphi}^2+\gamma \rho\right)\; .
\end{equation}

Following~\cite{CLW 98,Billyard+Coley+Hoogen 98} we introduce the new time
variable $N= \ln{\tilde{a}}$ and the expansion normalized
variables
\begin{eqnarray}
x &=& \frac{\aleph \dot{\varphi}}{\sqrt{6}H} \\
y &=& \frac{\aleph V(\varphi)}{\sqrt{3}H}
\end{eqnarray}
where $\aleph^2 = 8\pi G$ (in what follows we shall adopt units that set $\aleph^2 =1$).

We thus obtain the following third order dynamical system
\begin{eqnarray}
(x)' &=& -3x-\sqrt{\frac{3}{2}}\,
\left(\frac{\partial_{\varphi}V}{V}\right)\, y^2 +
\frac{3}{2}x\,\left[2x^2+\gamma\,
(1-x^2-y^2)\right] \label{x'}\\
y' &=& \sqrt{\frac{3}{2}}\, \left(\frac{\partial_\varphi
V}{V}\right)\, x y + \frac{3}{2}y\,\left[2
x^2+\gamma\,(1-x^2-y^2)\right] \; , \label{y'}\\
\varphi' &=& \sqrt{6} \, x \; . \label{vphi'}
\end{eqnarray}
where we have used
$\rho/3H^2=\Omega_m= 1-x^2
- y^2$.

The latter equation (\ref{vphi'}) is usually overlooked and only
the former two are considered in the previous work on scaling
solutions. This is due to the fact that if one considers an
exponential potential $V(\varphi) \propto \exp{(\lambda \varphi)}$
the two equations, (\ref{x'}) and (\ref{y'}), form a two
dimensional closed autonomous system of the polynomial type whose
qualitative behaviour can be easily determined, and which gives
rise to scaling (tracker) solutions~\cite{CLW 98,Billyard+Coley+Hoogen 98}.

Here, we shall continue to consider the full system.
We immediately see 
from the consideration of equations
%the complete system
(\ref{x'}-\ref{vphi'}) that they permit the
study of any other positive potentials $V(\varphi)$ besides the
exponential. Whenever $V(\varphi)\neq V_0 \,\exp{(\lambda
\varphi)}$ equation (\ref{vphi'}) tells us that all
singular points must have $x=0$. The former equations then
reduce to 
\begin{eqnarray}
x'&=& -\sqrt{\frac{3}{2}}\,\partial_\varphi \ln V\,y^2 \label{x':x=0}\\
y'&=& \frac{3}{2}\,\gamma\,y (1-y^2) \; , \label{y':x=0}
\end{eqnarray}

We may rule out the line of equilibria $(x=0,y=0,\varphi )$, which
corresponds to a fluid dominated universe and is unstable in the
$y$ direction with eigenvalue $3\gamma/2 \ge 0$. Scaling solutions,
in the sense of power law behaviour for the scalar field energy density,
for fluid dominated universes were studied by Liddle and Scherrer~\cite{Liddle+Scherrer 99}.
In this approximation, they found a late
time attractor for which both the kinetic and the potential energy
of the field behave as a certain power of the inverse scale factor,
while the dominant component scales as a smaller power of the
inverse scale factor, for potentials with null extrema.
Their analysis cannot be retrieved through the study of system (\ref{x'}-\ref{vphi'}),
which becomes singular when $V(\varphi )=0$. Anyway, our interest is
the search for trackers, that is, solutions for which the field's energy density (and its kinetic energy and potential as well) and the fluid's energy density
all scale as the same power of the scale factor. 
And this means equilibria of (\ref{x'}-\ref{vphi'}) in the region $0<x^2+y^2<1$.
Considering again the system (\ref{x'}-\ref{vphi'}), we see that there exists only another
class of fixed points, namely those of the form
\begin{equation}
p_{crit}=(x=0,y=1,\varphi=\varphi_0) \; ,
\end{equation}
where $\varphi_0$ is such that $\partial_\varphi V =0$.
This equilibrium point is degenerate, but it is easy to check
taking into account the second order terms in the $y'$ equation
that it is stable provided that  $\varphi_0$ is a minimum of $V(\varphi )$.
Again, this is no tracker solution. In this case, we have the opposite scenario,
a field dominated universe as a result  of an effective cosmological constant.

This analysis has lead many authors to claim that only exponential
potentials may give rise to tracker solutions~\cite{Liddle+Scherrer 99}. However, we must
still explore the possibility of the existence of equilibria at $\varphi =
\pm \infty$. In fact, the phase space of the full three dimensional system
(\ref{x'}-\ref{vphi'}) is not compact, and there may exist late time attractors
with the required behaviour on the boundary $\varphi =\pm \infty$ .
In what follows, we shall focus on a particular potential,
\begin{equation}
V(\varphi ) =
\left (\frac {2\Gamma }
{\alpha +\alpha c(\varphi )}
\right )^{1/(\lambda-1)}
\left (\frac {(2+\Gamma)+ (2-\Gamma) c(\varphi )}
{2( 1 + c(\varphi ))}
\right )\; , \label{V_b+p}
\end{equation}
where
\begin{equation}
c(\varphi ) = \cosh (\sqrt{3\Gamma} (\lambda -1)\varphi ), 
\end{equation}
and  $\Gamma$, $\alpha$ and $\lambda >1$ are positive constants. This particular potential is motivated by the fact
that, in the absence of any other matter component, it corresponds to an equation of state of the form
\begin{equation}
p + \rho = \Gamma \rho - \alpha \rho ^{\lambda }, \label{eqst_b+p}
\end{equation} 
which is a natural generalization of the 'exponential' equation of state
\begin{equation}
p + \rho = \gamma \rho . 
\end{equation} 
This is only to fix ideas, and it will become clear that the only essential
feature of the potential is that it should behave asymptotically as an
exponential potential. Barrow~\cite{Barrow 88} has derived   exact solutions of string-driven cosmologies introducing a bulk viscosity pressure which yields an equation of state equivalent to (\ref{eqst_b+p}) when the spatial curvature vanishes. 

Introducing the auxiliary variable $\psi = 1/\varphi $,
we obtain
\begin{equation}
\psi '= - \psi ^2\sqrt{6} \, x \ \label{psi'}
\end{equation}
which shows that the 'infinity manifold' $\psi =0$ is invariant. The flow
on this invariant manifold is given by the following equations
obtained from Eqs.~(\ref{x'},\ref{y'}) in the limit $\psi\to 0$
\begin{eqnarray}
(x)' &=& -3x+k\sqrt{\frac{3}{2}}\,
 y^2 +
\frac{3}{2}x\,\left[2x^2+\gamma\,
(1-x^2-y^2)\right] \label{x'inft}\\
y' &=& -k\sqrt{\frac{3}{2}}\,  x y + \frac{3}{2}y\,\left[2
x^2+\gamma\,(1-x^2-y^2)\right] \; , \label{y'inft}
\end{eqnarray}
where $k=\sqrt{3\Gamma}$. This system was studied in~\cite{CLW 98,Billyard+Coley+Hoogen 98} (note that in Ref.~\cite{CLW 98} the potential is $V=V_0\exp{-\lambda\varphi}$, i.e., $\lambda$ is used instead of $k$), and it has a sink whenever $k^2>3\gamma $ located at
$(x=\sqrt {3/2}\gamma /k, y=\sqrt {3(2-\gamma )\gamma /(2k^2)}$. Since
the $x$ coordinate for this sink is positive, equation (\ref{psi'}) tells
us that this equilibrium point is a sink for the full regularized system
on the $\varphi = +\infty $ invariant boundary.
Moreover, this equilibrium point is the only stable equilibrium of the
full system, and will therefore attract almost all the initial conditions.
Therefore, just as in the case of the exponential potential, the only
late time attractor for the system for these values of the parameters
is a tracker scaling solution. This conclusion carries over to every
positive and monotonous potential, provided
that its asymptotic behaviour is exponential. In fact for all these potentials the projection of the late-time flow on the $(x,y)$-plane is qualitatively similar to the flow of the autonomous system (\ref{x'inft},\ref{y'inft}) associated with the exponential potential.

\section{Discussion}
The main conclusion of our work is that tracker solutions are compatible with a wide class of potentials and thus are not restricted to the exponential potential case. 
The potentials must be positive and monotonous, and have to asymptotically approach the exponential behaviour. As the qualitative behaviour of the models based on the extended class of tracker potentials discussed in this paper is basically the same as that of the models based on exponential potentials, they  will face the same sort of difficulties as those faced by the latter~\cite{CLW 98,Ferreira+Joyce 97,Zlatev+Wang+Steinhardt 99,Liddle+Scherrer 99} in the later stages of the evolution of the universe.

However in the early universe the differences that are allowed with regard to the exponential potential permit greater flexibility in building models with  scalar fields. For instance the self-interacting scalar field cosmologies based on the potential of Eq.~(\ref{V_b+p}) have an early-time regime dominated by the scalar field (when $x,y\simeq 0$). During this stage the effective equation of state is $\gamma_\varphi \simeq \Gamma s/(s+\alpha)$, where $s=(a/a_0)^{3(\lambda-1)\Gamma}$ and thus, for $s$ sufficiently small, $\gamma_\varphi\simeq 0$ yielding slow-roll inflation. Indeed the relevant part of the potential during such a stage is
$V(\varphi)\simeq (\Gamma/\alpha)^{1/(\lambda-1)}\left(1-\nu\varphi^2\right)$, where $\nu=3\Gamma(\lambda-1)[\lambda-(2-\Gamma)(\lambda-1)]/8$ is a constant, and   is akin to the pseudo-Nambu-Golsdtone boson of natural inflation~\cite{Freese et al 90} which gives slow-roll without fine-tuning~\cite{Charters+Mim+Nunes 00}. The required minimum number of e-foldings is guaranteed by $3\Gamma(\lambda-1)\los \ln{\alpha}/60$, and hence does not restrict the value of $\Gamma$. Since the potential~(\ref{V_b+p}) eventually yields the tracker behaviour (when $\Gamma>\gamma_{matter}$), it permits a smooth transition out of quasi-de Sitter inflation into an epoch dominated by the matter. In particular, if $\Gamma>\gamma_{matter}=4/3$, the scalar field self-adjusts to the radiation behaviour and the scale factor evolves as a power-law $a(t)\propto t^{1/2}$. Such a smooth transition cannot be achieved with a bare exponential potential which either yields eternal inflation (whenever $k^2<3\gamma_{matter}$) or, alternatively, yields the tracking behaviour without any stage of inflation. In this latter case inflation is usually fulfilled through the additional consideration of yet another scalar field devoted to that purpose. 
The simpler model based on the potential~(\ref{V_b+p}) thus illustrates the underlying possibilities of the wider class of tracking potentials presented in this work to try and meet the various theoretical and observational constraints that have to be faced by a scalar field cosmology.

\section*{Acknowledgements}
The authors wish to acknowledge the finantial support from Funda\c c\~ao de Ci\^encia e Tecnologia  under the grant
PBIC/C/FIS/2215/95. 

%%%%%%%%%%%%%%%%%%%%%%%%%%%%%%%%%%%%%%%%%%%%%%
%\section*{References}

\end{document}